\documentclass{aa501}    
\usepackage{graphicx}
\begin{document}

\title{Chemical composition of red horizontal branch stars
in the thick disk of the Galaxy\thanks{Based on observations obtained at
the Nordic Optical Te\-les\-co\-pe, La Palma}\fnmsep\thanks{Table 1 is only
available in electronic at the CDS via anonymous ftp to
cdsarc.u-strasbg.fr (130.79.128.5) or via
http://cdsweb.u-strasbg.fr/Abstract.html} }

   \author{G. Tautvai\v{s}ien\.{e}\inst{1} \and B. Edvardsson\inst{2}
   \and I. Tuominen\inst{3} \and I. Ilyin \inst{3}}

   \offprints{G. Tautvai\v{s}ien\.{e},\\ 
   \email{taut@itpa.lt}}

   \institute{Institute of Theoretical Physics and Astronomy (ITPA),
              Go\v{s}tauto 12, Vilnius 2600, Lithuania
        \and  Uppsala Astronomical Observatory, Box 515,
	      SE-751\thinspace20 Uppsala, Sweden
	\and Astronomy Division, Department of Physical Sciences,
              P.O. Box 3000, 90014 University of Oulu, Finland
	      }

   \authorrunning{G. Tautvai\v{s}ien\.{e} et al.}
   \titlerunning{Chemical composition of RHB stars
in the thick disk of the Galaxy}

   \date{Received 00 January, 2001; Accepted 00 January, 2001}

  \abstract{High-resolution spectra of 13 core helium-burning stars in the
thick disk of the Galaxy have been obtained with the SOFIN spectrograph
on the Nordic Optical Telescope to investigate abundances of up to 22
chemical elements. Abundances of carbon were studied using the C$_2$
Swan (0,1) band head at 5635.5~{\AA}.  The wavelength interval
7980--8130~{\AA} with strong CN features was analysed in order to determine
nitrogen abundances and $^{12}\mathrm{C}/^{13}\mathrm{C}$  isotope ratios.
The oxygen abundances were determined from the [O\,{\sc i}] line at 6300~{\AA}.
Abundances in the investigated stars suggest that carbon is depleted by
about 0.3~dex, nitrogen is enhanced by more than 0.4~dex and oxygen is
unaltered.
The $^{12}\mathrm{C}/^{13}\mathrm{C}$
ratios are lowered and lie between values 3 and 7 which is in agreement
with ``cool bottom processing'' predictions (Boothroyd \&
Sackmann 1999). The C/N ratios in the investigated stars are lowered
to values between 0.7 and 1.2 which is less than present day theoretical
predictions and call for further studies of stellar mixing processes.
Abundance ratios of O, Mg, Eu and other heavy
chemical elements to iron in the investigated stars show a pattern
characteristic of thick disk stars. The results provide evidence that
the thick disk population has a distinct chemical history from the thin
disk. The onset of the bulk of SN Ia is suggested to appear at
[Fe/H]$\approx -0.6$~dex.
}


\maketitle

\section{Introduction}

``Upgren's Unclassified Stars: A New Type of G-Giant Stars ?''-- was the
title of a paper by Sturch \& Helfer (1971) in which $UBVRI$ photometry was
presented for 17 stars for which Upgren (1962) could not obtain luminosities.
Upgren (1962) had conducted objective-prism observations with dispersion
of 580~{\AA}/mm$^{-1}$ for late-type stars near the north galactic pole.
For the G stars, luminosity criteria were the two
CN bands at $\lambda 3820-80$~{\AA} and $\lambda 4130-4215$~{\AA}.
It appeared that for some G stars,
luminosity determination from these features was inaccurate.
Sturch \& Helfer, however, also met with difficulties: the position of stars
observed in the $U-B,R-I$ diagram matched neither the Hyades nor nearby field
dwarfs, nor field giants with $r<100$ pc, nor M$\thinspace 67$ or
the giant branches of a variety of globular clusters. The authors concluded that
these unclassified stars probably belong to the field equivalents of
the red horizontal branch (RHB) stars of metal-rich globular clusters.
This paper marked the beginning of a serious effort to study red horizontal
branch stars in the Galactic field (see Tautvai\v{s}ien\.{e} 1996a for a
review).

64 G stars from Upgren's list were investigated by Rose (1985) using a
quantitative three-dimensional spectral classification system employing
2.5~{\AA} resolution spectra in the blue. A number of Upgren's unclassified
stars were found to be dwarfs.
Quite a large group of Upgren's G stars were, however, shown to be evolved,
based on the strength of their Sr\,{\sc ii} $\lambda 4077$~{\AA} line.
They were also distinguished from
post-main-sequence stars evolving through the same region of the HR diagram
because of the unique appearance of their CN $\lambda 3883$ and 4216~{\AA}
bands. It was concluded that a class of red horizontal-branch stars,
similar to those in the ``metal-rich'' globular cluster M~71, has been
identified in the Galactic disk.
Moreover, it was noticed that these stars have metallicities and kinematics
which are common for
the `thick disk' of the Galaxy revealed by Gilmore \& Reid (1983).
Detailed measurements of kinematic parameters of
the stars by Stetson \& Aikman (1987) have confirmed that they belong to the
thick disk of the Galaxy.

Norris (1987) reported DDO observations for ten
of Upgren's red giants which Rose (1985) identified as RHB stars and
presented arguments that these stars could equally well be the
core-helium-burning `clump' stars similar to those seen in the old, metal
deficient open cluster NGC~2243. Consequently they could be as young as
about $5-7$ Gyr rather than about 14 Gyr as would follow from their identity
to the population of metal-rich galactic disc globular clusters.

Photometric observations and three-dimensional classification in the
Vilnius photometric system were carried out for 13 of the Roses's RHB
stars by Tautvai\v{s}ien\.{e} (1996b). The results were photometric spectral
types,
metallicities, effective temperatures, surface gravities, absolute magnitudes
and ages. The stars form a group with mean
[Fe/H]$=-0.6\pm0.1$ which is between $-0.7$ as evaluated by Rose (1985) and
$-0.5$ as determined by Norris (1987) from the cyanogen excess parameter
$\delta_{\rm CN}$. An age of about 10--12 Gyr was ascribed to the group
from comparison with model isochrones. This age is intermediate between the
ages of the disk globular clusters and the oldest open clusters.

The aim of the present study is to perform a high resolution spectroscopic
analysis of 13 red horizontal branch stars which were identified in
Upgren's list by Rose (1985).
We expect that C/N and $^{12}$C/$^{13}$C abundance ratios, and possibly also the
abundances of sodium, aluminium and $s$-process elements, will
provide information on the extent of mixing processes in these evolved stars.
Abundances of other chemical elements will be useful for the
interpretation of the chemical evolution of the thick disk of the Galaxy.

\section{Observations and data reductions}
  The spectra were obtained at the Nordic Optical Telescope (NOT) with the
SOFIN \'{e}chelle spectrograph (Tuominen 1992) in 1997.
The 2nd optical camera ($R\approx 60\thinspace 000$) and the 3rd optical camera
($R\approx 30\thinspace 000$) were used. The spectra were recorded with the
Astromed-3200 CCD camera (Mackay 1986)
equipped with an EEV P88100 UV-coated CCD of
$1152\times298$ pixels operated at the working temperature of 150 K.
With the 2nd camera we observed simultaneously 13 spectral orders,
each of 40--60~{\AA} in length, located from 5650~{\AA} to 8130~{\AA} and
with the 3rd camera we observed 25 spectral orders,
each of 80--150~{\AA} in length, located from 4500~{\AA} to 8750~{\AA}.
All spectra were exposed to S/N$\ge100$.

Reductions of the CCD images were made with the {\it 3A} software package
(Ilyin 1996, 2000). Procedures of bias subtraction, spike elimination, flat 
field
correction, scattered light subtraction, extraction of spectral orders were
used for image processing. A Th-Ar comparison spectrum was used for the
wavelength calibration. The continuum was defined by a number of narrow
spectral regions, selected to be free of lines in the solar spectrum.

The lines suitable for measurement were chosen using the requirement that the
profiles be sufficiently clean to provide reliable equivalent widths.
Inspection of the solar spectrum (Kurucz et al.\ 1984) and
the solar line identifications of Moore et al.\ (1966) were used to avoid
blends. Lines blended by telluric absorption lines were omitted from
treatment as well. The equivalent widths of lines were measured by fitting
of a Gaussian function.
The line measurements are listed in Table~1 (available in electronic
form at CDS).

\section{Method of analysis }

The spectra were analysed using a differential model atmosphere technique.
The method of analysis and atomic line parameters are the same as used
recently by Tautvai\v{s}ien\.{e} et al.\ (2000, Paper~I), where the chemical
composition of evolved stars in the open cluster M~67 was investigated.
The {\it Eqwidth} and {\it Spectrum} programme packages, developed at
Uppsala Astronomical Observatory, were used to carry out the
calculations of abundances from measured equivalent widths and synthetic
spectra, respectively. A set of
plane parallel, line-blanketed, flux constant LTE model atmospheres with 
solar abundance ratios was
computed by M. Asplund (Uppsala Astronomical Observatory)
with the updated version of the MARCS code (Gustafsson et al.\ 1975)
using continuous opacities from Asplund et al.\ (1997) and including UV line
blanketing as described by Edvardsson et al. (1993).
The solar model atmosphere for the differential analysis was also
calculated in Uppsala (Edvardsson et al.\ 1993).

The effective temperatures for the programme stars were initially taken from
Tautvai\v{s}ien\.{e} (1996b), where they were derived using the
intrinsic colour index $(Y-V)_0$ of the Vilnius photometric system.
For the star BD +28\degr 2079 the $(Y-V)_0$ was taken from 
Bartkevi\v{c}ius \& Lazauskait\.{e} (1997) and the same procedure applied. 
In the work by Tautvai\v{s}ien\.{e} (1996b) as well as in the work by Norris 
(1987) the interstellar reddening for these stars was accepted to be zero. 
Although these stars are located in the direction of the North Galactic Pole, 
where the reddening should be small, Bartkevi\v{c}ius \& Lazauskait\.{e} (1997) 
have found that some of the stars are affected. We decided to introduce a 
spectroscopic method to solve the problem. We corrected, when needed, the 
effective temperatures by achieving the LTE exitation balance in the iron 
abundance results. For nine stars the effective temperatures were adjusted by 
$-70$ -- $+110$\,K.

The surface gravities were found by forcing  Fe\,{\sc i} and Fe\,{\sc ii} to
yield the same iron abundances, 47 Fe\,{\sc i} and 5 Fe\,{\sc ii} lines were 
used. The microturbulent velocities were determined  by
forcing Fe\,{\sc i} line abundances to be independent of the equivalent width.
The derived atmospheric parameters are listed in Table~2. 

Abundances of carbon and nitrogen were determined using the
spectrum synthesis technique.
The interval of 5632--5636~{\AA} was synthesized and
compared with observations in the vicinity of the ${\rm C}_2$ Swan 0--1 band
head at 5635.5~{\AA}.
The 5635.5~{\AA} ${\rm C}_2$ band head is strong enough in our
spectra and is quite sensitive to changes of the carbon abundance (see
Fig.~1 for illustration).
The same atomic data of ${\rm C}_2$ as used by
Gonzalez et al.\ (1998) and in Paper~I were adopted for the analysis.

\begin{figure}
\resizebox{\hsize}{!}{\includegraphics{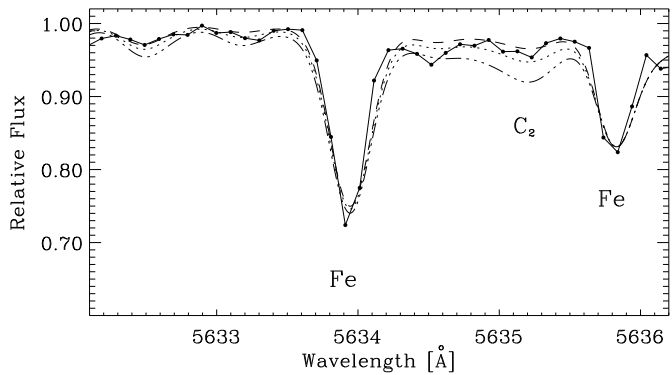}}
\caption{Synthetic (dashed and dotted curves) and observed
(solid curve with dots) spectra for the 1--0 C$_2$ region near
$\lambda 5635$~{\AA} of BD +34\degr 2371.
The syntheses were generated with [C/H]$=-0.3$, $-0.4$, and $-0.5$ 
(dashed-dotted, dotted and dashed curves, respectively)} 
\label{GT1664f1}
\end{figure}

The intervals of 7980--8130~{\AA} with $R\approx 30\,000$ and
8380--8430~{\AA} with $R\approx 60\,000$, containing strong CN
features, were analysed in order to determine the nitrogen abundance.
The $^{12}$C/$^{13}$C determination was based on the 8004.728~{\AA}
$^{13}$CN feature. 11 other weaker $^{13}$CN features
($\lambda$ 7989.45, 8010.4, 8011.2, 8016.35, 8022.65, 8036.15, 8043.2,
8048.3, 8051.8, 8056.4, 8058.2 and 8065.0~{\AA}) were used
for error estimation. The molecular data for
$^{12}$C$^{14}$N and $^{13}$C$^{14}$N
were taken from {\it ab initio} calculations of CN isotopic line strengths,
energy levels and wavelengths by Plez (1999), with all $gf$ values
increased
by $+0.03$~dex in order to fit our model spectrum to the solar atlas of
Kurucz et al.\ (1984).
The $^{13}$CN line wavelengths were, however, adopted from laboratory
measurements by Wyller (1966).
Parameters of atomic lines in the spectral synthesis intervals were adopted
from the VALD database (Piskunov et al.\
1995). In order to check the correctness of the
input data, synthetic spectra of the Sun were compared to the
solar atlas of Kurucz et al.\ (1984) and necessary adjustments were made
to the line data.

Fig.~2 illustrates the enhancement of the $^{13}$CN line at
8004.7~{\AA} in a spectrum of the star BD +27\degr 2057.

\begin{figure}
    \resizebox{\hsize}{!}{\includegraphics{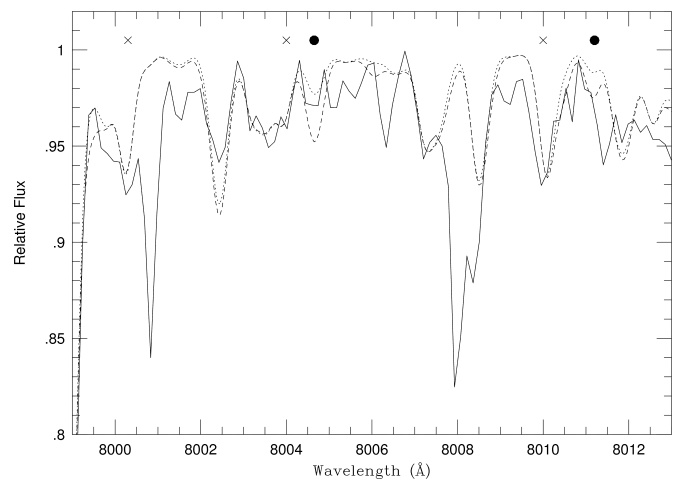}}
    \caption{
    A small portion of the 8000~{\AA} wavelength interval
    showing the 8004.7~{\AA}  $^{13}\mathrm{CN}$ feature in the star
    BD +27\degr 2057.
   The dotted line shows a synthetic spectrum
    with [N/H]=$-$0.28 and $^{12}\mathrm{C}/^{13}\mathrm{C}=7$, the
    dashed line shows a synthetic spectrum with [N/H]=$-$0.22 and
    $^{12}\mathrm{C}/^{13}\mathrm{C}=3$. The dots indicate features
    dominated by $^{13}\mathrm{CN}$, and the crosses mark features
    dominated by $^{12}\mathrm{CN}$. Unfitted features belong to the 
    Earth atmosphere }
    \label{GT1664f2}
\end{figure}


Abundances of oxygen were determined using equivalent widths of the 
[O\,{\sc i}] forbidden line at 6300~{\AA}, widely used in analyses of 
other late-type stars. This line was recently reexamined in the solar 
spectrum with a three-dimensional time-dependent hydrodynamical model 
solar atmosphere and implications of the Ni\,{\sc i} 
blend on oxygen abundances discussed (Prieto et al.\ 2001). Our test 
calculations showed that in our sample of stars the influence of the 
Ni line is very small (oxygen abundance changes do not exceed 
0.01--0.03~dex).      

The interval of 6643--6648~{\AA}, containing the Eu\,{\sc ii} line at 6645~{\AA}, was
computed in order to determine the europium abundance
 (see Fig.~3 for illustration).
The oscillator strength of the Eu\,{\sc ii} line, $\log gf=0.17$,
was adopted from Gurtovenko \& Kostik (1989). The solar abundance of
europium, later used for the differential analysis,
$\log A(\mathrm{Eu})\sun=0.49$, was determined
by fitting of the Kurucz et al. (1984) solar flux spectrum.
Parameters of other lines in the interval
were compiled from the VALD database. CN lines were also included, but none
of them seems to affect the europium line significantly.

\begin{figure}
\resizebox{\hsize}{!}{\includegraphics{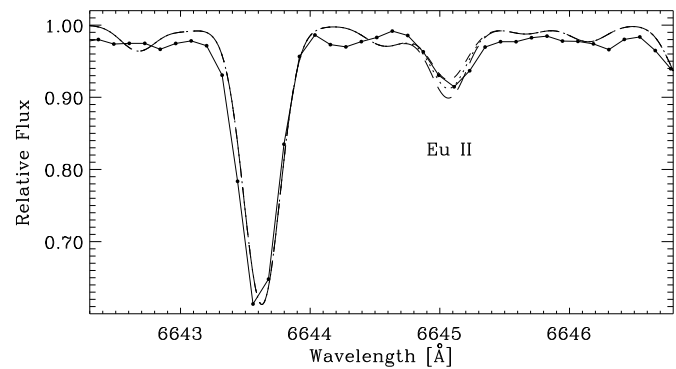}}
\caption{Synthetic and observed (thick solid curve and dots)
spectra for the region around the Eu\,{\sc ii} line at $\lambda$ 6645~{\AA}
in BD +27\degr 2057. The syntheses are generated with [Eu/Fe]=0.4,
0.5 and 0.6 (dashed, dotted and long-dashed curves, respectively)}
 \label{GT1664f3}
\end{figure}

   \begin{table}
  \setcounter{table}{1}
\centering
      \caption{Atmospheric parameters derived for the field RHB stars. The
last two columns give numbers of spectra observed with the resolving power
$R{\rm 1}\approx 30\, 000$ and $R{\rm 2}\approx 60\, 000$ }
      \[
         \begin{tabular}{lccrccc}
            \hline
            \noalign{\smallskip}
 BD/HD  & $T_{\rm e},{\rm K}$ &
            $\log g$ & [Fe/H] &
            ${ v_{\rm t},{\rm km/s}}$&$R{\rm 1}$&$R{\rm 2}$\\
            \noalign{\smallskip}
            \hline
            \noalign{\smallskip}
 \object{+25\degr 2436} & 4990 & 2.4 & $-$0.48 & 1.7 & 2 & 2 \\
 \object{+25\degr 2459} & 4980 & 2.5 & $-$0.35 & 1.5 & 1 & 2 \\
 \object{+25\degr 2502} & 5090 & 2.2 & $-$0.74 & 1.3 &   & 2 \\
 \object{+27\degr 2057} & 4840 & 2.1 & $-$0.60 & 1.7 & 1 & 2 \\
 \object{+28\degr 2079} & 4950 & 2.5 & $-$0.44 & 2.0 & 2 &   \\
 \object{+29\degr 2231} & 5060 & 2.5 & $-$0.39 & 1.9 & 2 & 2 \\
 \object{+29\degr 2294} & 5020 & 2.1 & $-$0.54 & 1.7 & 1 &   \\
 \object{+29\degr 2321} & 4980 & 2.3 & $-$0.50 & 1.9 & 2 &   \\
 \object{+33\degr 2280} & 5000 & 2.4 & $-$0.48 & 1.6 &   & 1 \\
 \object{+34\degr 2371} & 4980 & 2.5 & $-$0.18 & 1.6 & 1 &   \\
 \object{+36\degr 2303} & 4700 & 1.8 & $-$0.76 & 2.0 & 2 &   \\
 \object{104783}        & 5140 & 2.4 & $-$0.55 & 1.5 & 1 & 3 \\
 \object{105944}        & 5090 & 2.1 & $-$0.37 & 1.4 & 2 &   \\
             \noalign{\smallskip}
            \hline
         \end{tabular}
      \]
   \end{table}


Typical internal error estimates for the atmospheric parameters are:
$\pm~100$~K for $T_{\rm eff}$, $\pm 0.3$~dex for $\log g$ and
$\pm 0.3~{\rm km~s}^{-1}$ for $v_{\rm t}$. The sensitivity of the abundance
estimates to changes in the atmospheric parameters by the assumed errors is
illustrated  for the star BD +25\degr 2436 (Table~3). It is
seen that our estimated parameter uncertainties
do not affect the abundances seriously; the
element-to-iron ratios, which we use in our discussion, are even less
sensitive. The small differences between the chemical composition of the models and
the final abundance results have a neglible effect on the results.  
The $^{12}{\rm C}/^{13}{\rm C}$ ratios are not sensitive to the
model parameters or errors in the $\log gf$ values since they are determined after
fitting the $^{12}{\rm CN}$ features.

The scatter of the deduced line abundances $\sigma$, presented in 
Table~4,
gives an estimate of the uncertainty coming from the random errors in
the line parameters (e.g.\ random errors in equivalent widths, oscillator
strengths and possible undetected line blends).
The approximate value of these uncertainties amounts in the mean
to $\sigma$=$0.10$~dex.
Other sources of observational errors, such as continuum placement or
background subtraction problems are partly included in the equivalent width
uncertainties.
The nitrogen abundance is less dependent on line
measurement uncertainties because, depending on the number of spectra
observed, the number of CN lines used for the analysis was ranging from 34
to 162.

Since the abundances of C, N and O are tied together by the molecular equilibria
in the stellar atmospheres and the abundances were determined in the sequence
O (from [O\,{\sc i}]) $\Rightarrow$ C (from C$_2$) $\Rightarrow$ N (from CN),
we have investigated how an error in one of them would typically affect
our abundance determinations of the others. Calculations for BD +25\degr 2436:
a change of the oxygen abundance of
$\Delta{\rm [O/H]}=-0.10$ would result in
$\Delta{\rm [C/H]}=-0.04$,
$\Delta{\rm [N/H]}=-0.01$ and thus
$\Delta{\rm [C/N]}=-0.03$;
a change in
$\Delta{\rm [C/H]}=-0.10$ would cause
$\Delta{\rm [N/H]}=+0.10$,
$\Delta{\rm [C/N]}=-0.20$ and
$\Delta{\rm [O/H]}=-0.03$;
$\Delta{\rm [N/H]}=-0.10$ has no appreciable effect on either the oxygen
or the carbon molecular equilibria (except for CN).
Note in particular that the C/N ratios are sensitive to the carbon
abundance uncertainties squared.

   \begin{table}
   \setcounter{table}{2}
   \centering
      \caption{Effects on derived abundances resulting from model changes
for the star BD\thinspace +25\degr 2436. The table entries show the
effects on the
logarithmic abundances relative to hydrogen, $\Delta$[A/H]. Note that the
effects on ``relative'' abundances, for example [A/Fe], are often
considerably smaller than abundances relative to hydrogen, [A/H] }
      \[
         \begin{tabular}{lrrr}
            \hline
            \noalign{\smallskip}
 Ion & ${ \Delta T_{\rm eff} }\atop{ -100 {\rm~K} }$ &
            ${ \Delta \log g }\atop{ -0.3 }$ &
            ${ \Delta v_{\rm t} }\atop{ -0.3 {\rm km~s}^{-1}}$ \\
            \noalign{\smallskip}
            \hline
            \noalign{\smallskip}
   C (C$_2$)   &  0.02  &$-$0.03 &  0.00 \\
   N (CN)       &$-$0.10 &$-$0.03 &  0.00  \\
   O\,{\sc i}   &$-$0.01 &$-$0.13 &  0.00  \\
   Na\,{\sc i}  &$-$0.07 &  0.01  &$-$0.05  \\
   Mg\,{\sc i}  &$-$0.04 &$-$0.01 &$-$0.03  \\
   Al\,{\sc i}  &$-$0.05 &  0.01  &$-$0.02  \\
   Si\,{\sc i}  &  0.01  &$-$0.04 &  0.03  \\
   Ca\,{\sc i}  &$-$0.10 &  0.01  &$-$0.11  \\
   Sc\,{\sc i}  &$-$0.12 &  0.00  &  0.02  \\
   Sc\,{\sc ii} &  0.02  &$-$0.13 &  0.10  \\
   Ti\,{\sc i}  &$-$0.14 &  0.01  &  0.09  \\
   Ti\,{\sc ii} &  0.01  &$-$0.12 &  0.08  \\
   V\,{\sc i}   &$-$0.16 &  0.00  &  0.03  \\
   Cr\,{\sc i}  &$-$0.11 &  0.01  &$-$0.09  \\
   Mn\,{\sc i}  &$-$0.08 &$-$0.01 &  0.04  \\
   Fe\,{\sc i}  &$-$0.08 &$-$0.02 &  0.06  \\
   Fe\,{\sc ii} &  0.09  &$-$0.14 &  0.10  \\
   Co\,{\sc i}  &$-$0.08 &$-$0.02 &$-$0.02  \\
   Ni\,{\sc i}  &$-$0.05 &$-$0.03 &  0.08  \\
   Y\,{\sc i}   &$-$0.17 &$-$0.01 &  0.02  \\
   Y\,{\sc ii}  &  0.00  &$-$0.14 &  0.13  \\
   Zr\,{\sc i}  &$-$0.17 &  0.00  &$-$0.01  \\
   Ba\,{\sc ii} &$-$0.02 &$-$0.11 &  0.27  \\
   La\,{\sc ii} &$-$0.01 &$-$0.13 &  0.01  \\
   Sm\,{\sc ii} &$-$0.02 &$-$0.14 &  0.03  \\
   Eu\,{\sc ii} &  0.00  &$-$0.10 &$-$0.01  \\
                 \noalign{\smallskip}
            \hline
         \end{tabular}
      \]
   \end{table}

\section{Relative abundances}

The abundances relative to hydrogen
[A/H]\footnote{In this paper we use the customary spectroscopic notation
[X/Y]$\equiv \log_{10}(N_{\rm X}/N_{\rm Y})_{\rm star} -
\log_{10}(N_{\rm X}/N_{\rm Y})_\odot$} and $\sigma$ (the line-to-line
scatter) derived for up to 26 neutral and ionized species for the programme
stars are listed in Table~4.
The abundances of barium are corrected for
non-LTE effects by the subtraction of 0.20~dex (see Subsec.~4.5 for discussion).

   \begin{table*}
  \setcounter{table}{3}
\centering
      \caption{Abundances relative to hydrogen [A/H] derived
for programme stars. The quoted
errors, $\sigma$, are the standard deviations in the mean value due to the
line-to-line scatter within the species. Details on error estimates of the
$^{12}$C/$^{13}$C ratios are described in \$ 4.1.
The number of lines used is indicated by $n$. }
      \[
         \begin{tabular}{lrcrcrcrcrcrcrcrc}
            \hline
            \noalign{\smallskip}
  & \multicolumn{3}{c}{BD +25\degr 2436} &
  & \multicolumn{3}{c}{BD +25\degr 2459} &
  & \multicolumn{3}{c}{BD +25\degr 2502} &
  & \multicolumn{3}{c}{BD +27\degr 2057} \\
            \noalign{\smallskip}
\cline{2-4}\cline{6-8}\cline{10-12}\cline{14-16}
            \noalign{\smallskip}
Ion &[A/H] &$\sigma$ &$n$&\ &[A/H] &$\sigma$ &$n$&\ &[A/H]&$\sigma$&$n$
&\ &[A/H] &$\sigma$ &$n$\\
            \noalign{\smallskip}
            \hline
            \noalign{\smallskip}
C (C$_2$)   &$-$0.63&    &1  & &$-$0.58&    &1  & &      &    &1 & &$-$0.78&    &1  \\
N (CN)      &$-$0.01&0.14&154& &  0.14&0.11&111& &      &    &  & &$-$0.27&0.15&102\\
O\,{\sc i}  &$-$0.25&    &1  & &$-$0.24&    &1  & &$-$0.24&    &1 & &$-$0.25&    &1  \\
Na\,{\sc i} &$-$0.54&0.01&2  & &$-$0.28&0.15&2  & &$-$0.72&0.02&2 & &$-$0.40&0.10&2  \\
Mg\,{\sc i} &$-$0.21&0.04&2  & &$-$0.24&0.09&2  & &$-$0.32&    &1 & &$-$0.20&0.10&2  \\
Al\,{\sc i} &$-$0.47&0.06&4  & &$-$0.34&0.08&4  & &$-$0.68&0.05&3 & &$-$0.45&0.03&4  \\
Si\,{\sc i} &$-$0.31&0.12&14 & &$-$0.18&0.12&13 & &$-$0.48&0.10&6 & &$-$0.24&0.11&14 \\
Ca\,{\sc i} &$-$0.36&0.12&7  & &$-$0.23&0.12&8  & &$-$0.53&0.14&4 & &$-$0.39&0.15&7  \\
Sc\,{\sc i} &$-$0.40&0.10&4  & &$-$0.31&0.11&4  & &$-$0.39&    &1 & &$-$0.56&0.15&4  \\
Sc\,{\sc ii}&$-$0.33&0.11&10 & &$-$0.20&0.08&11 & &$-$0.43&0.16&6 & &$-$0.46&0.11&10 \\
Ti\,{\sc i} &$-$0.29&0.12&23 & &$-$0.11&0.15&21 & &$-$0.57&0.16&6 & &$-$0.32&0.13&23 \\
Ti\,{\sc ii}&$-$0.26&    &1  & &$-$0.05&    &1  & &      &    &  & &$-$0.33&    &1  \\
V\,{\sc i}  &$-$0.40&0.10&17 & &$-$0.22&0.12&18 & &$-$0.70&0.10&6 & &$-$0.50&0.15&18 \\
Cr\,{\sc i} &$-$0.38&0.04&7  & &$-$0.23&0.13&7  & &      &    &  & &$-$0.55&0.09&7  \\
Mn\,{\sc i} &$-$0.46&0.08&3  & &$-$0.31&0.09&3  & &      &    &  & &$-$0.70&0.08&2  \\
Fe\,{\sc i} &$-$0.48&0.12&43 & &$-$0.35&0.10&42 & &$-$0.74&0.06&18 & &$-$0.60&0.12&40\\
Fe\,{\sc ii}&$-$0.48&0.10&5  & &$-$0.35&0.13&5  & &$-$0.74&0.11&2 & &$-$0.60&0.12&5  \\
Co\,{\sc i} &$-$0.41&0.13&10 & &$-$0.30&0.14&9  & &$-$0.58&0.04&2 & &$-$0.47&0.14&8  \\
Ni\,{\sc i} &$-$0.42&0.14&22 & &$-$0.24&0.13&20 & &$-$0.80&0.09&8 & &$-$0.56&0.15&21 \\
Y\,{\sc i}  &$-$0.37&    &1  & &$-$0.28&    &1  & &      &    &  & &$-$0.53&    &1  \\
Y\,{\sc ii} &$-$0.44&0.03&3  & &$-$0.29&0.06&3  & &      &    &  & &$-$0.46&0.11&2  \\
Zr\,{\sc i} &$-$0.50&0.11&3  & &$-$0.32&0.13&4  & &$-$0.41&0.06&3 & &$-$0.67&0.13&4  \\
Ba\,{\sc ii}&$-$0.70&0.07&2  & &$-$0.34&0.04&2  & &$-$0.69&0.05&2 & &$-$0.82&0.03&2  \\
La\,{\sc ii}&$-$0.62&    &1  & &$-$0.38&    &1  & &      &    &  & &$-$0.68&    &1  \\
Sm\,{\sc ii}&$-$0.53&    &1  & &$-$0.05&    &1  & &      &    &  & &$-$0.36&    &1  \\
Eu\,{\sc ii}&$-$0.05&    &1  & &$-$0.05&    &1  & &      &    &  & &$-$0.15&    &1  \\
     &      &    &   & &           &   & &           &  & &      &    &   \\
C/N  &  0.96&    &   & &  0.76&    &   & &      &    &  & & 1.23 &   &    \\
$^{12}$C/$^{13}$C &  5 &+5/-2& & & 7 & +3/-2& & & &  &   & &    5 &+2/-2  &   \\
             \noalign{\smallskip}
   \cline{1-16}
         \end{tabular}
      \]
   \end{table*}

\subsection{Carbon and nitrogen}

   \begin{table*}
  \setcounter{table}{3}
\centering
      \caption{(continued)}
      \[
         \begin{tabular}{lrcrcrcrcrcrcrcrc}
            \hline
            \noalign{\smallskip}
  & \multicolumn{3}{c}{BD +28\degr 2079} &
  & \multicolumn{3}{c}{BD +29\degr 2231} &
  & \multicolumn{3}{c}{BD +29\degr 2294} &
  & \multicolumn{3}{c}{BD +29\degr 2321} \\
            \noalign{\smallskip}
\cline{2-4}\cline{6-8}\cline{10-12}\cline{14-16}
            \noalign{\smallskip}
Ion &[A/H] &$\sigma$ &$n$&\ &[A/H] &$\sigma$ &$n$&\ &[A/H]&$\sigma$&$n$
&\ &[A/H] &$\sigma$ &$n$\\
            \noalign{\smallskip}
            \hline
            \noalign{\smallskip}
C (C$_2$)    &       &    &  & &$-$0.60&    &1  & &$-$0.80&    &1  & &$-$0.68&    &1 \\
N (CN)       &       &    &  & &  0.15 &0.12&162& &$-$0.10&0.14& 34& &  0.00 &0.13&93\\
O\,{\sc i}   &$-$0.10&    &1 & &$-$0.22&    &1  & &$-$0.31&    &1  & &       &    &  \\
Na\,{\sc i}  &$-$0.51&    &1 & &$-$0.34&0.09&2  & &$-$0.28&    &1  & &$-$0.35&    &1 \\
Mg\,{\sc i}  &$-$0.21&    &1 & &$-$0.18&0.02&2  & &$-$0.14&    &1  & &$-$0.26&    &1 \\
Al\,{\sc i}  &$-$0.22&0.03&2 & &$-$0.32&0.08&4  & &$-$0.46&0.06&2  & &$-$0.29&0.12&2 \\
Si\,{\sc i}  &$-$0.20&0.12&6 & &$-$0.20&0.07&14 & &$-$0.30&0.08&8  & &$-$0.27&0.08&8 \\
Ca\,{\sc i}  &$-$0.22&0.13&5 & &$-$0.29&0.17&8  & &$-$0.39&0.11&5  & &$-$0.41&0.14&7 \\
Sc\,{\sc i}  &       &    &  & &$-$0.26&0.10&4  & &$-$0.48&0.07&3  & &$-$0.50&0.15&3 \\
Sc\,{\sc ii} &$-$0.19&0.07&8 & &$-$0.23&0.09&10 & &$-$0.32&0.10&9  & &$-$0.33&0.08&8 \\
Ti\,{\sc i}  &$-$0.12&0.17&16& &$-$0.18&0.12&22 & &$-$0.27&0.13&19 & &$-$0.43&0.14&20\\
Ti\,{\sc ii} &$-$0.14&    &1 & &$-$0.30&    &1  & &$-$0.25&    &1  & &$-$0.42&    &1 \\
V\,{\sc i}   &$-$0.17&0.16&12& &$-$0.18&0.12&17 & &$-$0.43&0.14&12 & &$-$0.49&0.13&14\\
Cr\,{\sc i}  &$-$0.27&0.13&7 & &$-$0.39&0.15&8  & &$-$0.51&0.13&7  & &$-$0.51&0.15&7 \\
Mn\,{\sc i}  &$-$0.32&0.12&2 & &$-$0.32&0.16&3  & &$-$0.70&0.05&2  & &$-$0.57&0.08&3 \\
Fe\,{\sc i}  &$-$0.44&0.06&25& &$-$0.39&0.12&43 & &$-$0.54&0.10&24 & &$-$0.50&0.08&27\\
Fe\,{\sc ii} &$-$0.44&0.12& 3& &$-$0.39&0.07&5  & &$-$0.54&0.12&3  & &$-$0.50&0.08&3 \\
Co\,{\sc i}  &$-$0.35&0.17&4 & &$-$0.28&0.16&9  & &$-$0.37&0.12&6  & &$-$0.41&0.08&7 \\
Ni\,{\sc i}  &$-$0.38&0.10&15& &$-$0.33&0.15&22 & &$-$0.45&0.11&21 & &$-$0.50&0.12&21\\
Y\,{\sc i}   &       &    &  & &$-$0.41&    &1  & &$-$0.53&    &1  & &$-$0.57&    &1 \\
Y\,{\sc ii}  &$-$0.43&0.14&2 & &$-$0.45&0.11&4  & &$-$0.64&0.01&2  & &$-$0.56&0.13&4 \\
Zr\,{\sc i}  &$-$0.42&0.04&2 & &$-$0.29&0.02&3  & &$-$0.42&0.15&4  & &$-$0.53&0.17&3 \\
Ba\,{\sc ii} &$-$0.36&0.03&2 & &$-$0.54&0.01&2  & &$-$0.67&    &1  & &$-$0.60&    &1 \\
La\,{\sc ii} &       &    &  & &$-$0.47&    &1  & &       &    &   & &       &    &  \\
Sm\,{\sc ii} &$-$0.06&    &1 & &$-$0.20&    &1  & &       &    &   & &-0.25  &    &1 \\
Eu\,{\sc ii} &       &    &  & &$-$0.05&    &1  & &$-$0.28&    &1  & &$-$0.22&    &1 \\
             &       &    &  & &       &    &   & &       &    &   & &       &    &  \\
C/N          &       &    &  & &  0.71 &    &   & &  0.79 &    &   & &  0.83 &    &  \\
$^{12}$C/$^{13}$C&   &    &  & &3.5    &+3/-1.5& & & 3    &+2/-1&  & & 4     &+4/-1 \\
   \noalign{\smallskip}
   \cline{1-16}
         \end{tabular}
\]
\end{table*}

The carbon abundances obtained in our work were compared with carbon
abundances  determined for dwarf stars
in the galactic disk. Gustafsson et al.\ (1999), using the forbidden
[C\,{\sc i}] line, performed an abundance analysis of carbon in a sample of 80
late F and early G type dwarfs.
Since carbon abundances obtained using
the [C\,{\sc i}] 8727~{\AA} line and ${\rm C}_2$ molecular lines are usually
consistent (c.f. Clegg et al.\ 1981), we expect no systematic shift to
be present because of the different abundance indicators used.
As is seen from Fig.~4, the ratios of [C/Fe] in our stars lie much below
the trend obtained for dwarf stars in the Galactic disk
(Gustafsson et al.\ 1999).

Abundances in the investigated stars suggest that carbon is depleted by
about 0.3~dex and nitrogen is enhanced by more than 0.4~dex.
These abundance alterations
of carbon and nitrogen are larger than those we obtained for the clump stars
in the old, solar-metallicity open cluster M~67 (Paper~I), but
smaller than was found for more metal deficient RHB stars by Gratton et
al.\ (2000a). This brings additional evidence that mixing processes are
metallicity dependent. The C/N ratios in the investigated
stars are lowered to values in the range 0.7 to 1.2 which is less than predicted
by present day stellar evolution calculations. Gratton et al. (2000a) receive even 
smaller C/N ratios for the two red horizontal branch stars with [Fe/H] about $-1.5$~dex. 
The $^{12}\mathrm{C}/^{13}\mathrm{C}$ ratios are lowered and lie between
values 3 and 7 which indicate extra-mixing processes to be quite strong. 
Six more metal-deficient RHB stars investigated by Gratton et al.\ (2000a) show 
$^{12}\mathrm{C}/^{13}\mathrm{C}$ ratios from 6 to 12.   
 
  \begin{figure}
    \resizebox{\hsize}{!}{\includegraphics{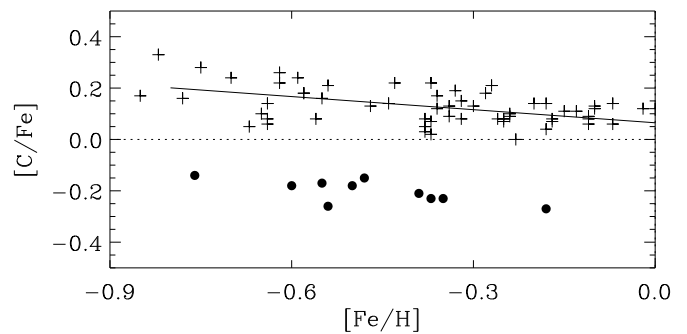}}
    \caption{[C/Fe] as a function of [Fe/H].
Results of this paper are indicated by {\it filled circles},
results obtained for
    dwarf stars of the galactic disk (Gustafsson et al.\ 1999) are indicated
by `{\it plus}' signs and the solid line. The relative underabundance in
the He-core burning stars is clearly seen }
\label{GT1664f4}
  \end{figure}


   \begin{table*}
  \setcounter{table}{3}
\centering
      \caption{(continued)}
      \[
         \begin{tabular}{lrcrcrcrcrcrcrcrc}
            \hline
            \noalign{\smallskip}
  & \multicolumn{3}{c}{BD +33\degr 2280} &
  & \multicolumn{3}{c}{BD +34\degr 2371} &
  & \multicolumn{3}{c}{BD +36\degr 2303} &
  & \multicolumn{3}{c}{HD 104783}\\
            \noalign{\smallskip}
\cline{2-4}\cline{6-8}\cline{10-12}\cline{14-16}
            \noalign{\smallskip}
Ion & [A/H] &$\sigma$ &$n$&\ &[A/H] &$\sigma$ &$n$&\ &[A/H]&$\sigma$&$n$& &[A/H]&$\sigma$&$n$\\
            \noalign{\smallskip}
            \hline
            \noalign{\smallskip}
C (C$_2$)    &       &    &  & &$-$0.45&    &1 & &$-$0.90&    &1 & &$-$0.72&    &1  \\
N (CN)       &       &    &  & &  0.13 &0.13&38& &$-$0.34&0.12&92& &$-$0.05&0.14&109\\
O\,{\sc i}   &$-$0.29&    &1 & &$-$0.13&    &1 & &       &    &  & &$-$0.14&    &1  \\
Na\,{\sc i}  &$-$0.33&0.02&2 & &  0.02 &    &1 & &$-$0.72&    &1 & &$-$0.56&0.11&2  \\
Mg\,{\sc i}  &$-$0.28&    &1 & &$-$0.05&    &1 & &$-$0.36&    &1 & &$-$0.23&0.06&2  \\
Al\,{\sc i}  &$-$0.41&0.04&4 & &$-$0.10&0.07&2 & &$-$0.55&0.05&2 & &$-$0.54&0.04&4  \\
Si\,{\sc i}  &$-$0.29&0.09&7 & &$-$0.06&0.06&7 & &$-$0.37&0.12&8 & &$-$0.24&0.08&15 \\
Ca\,{\sc i}  &$-$0.47&0.19&3 & &$-$0.07&0.15&7 & &$-$0.55&0.13&7 & &$-$0.34&0.13&7  \\
Sc\,{\sc i}  &$-$0.34&    &1 & &$-$0.10&    &1 & &$-$0.76&0.03&2 & &$-$0.56&0.09&2  \\
Sc\,{\sc ii} &$-$0.33&0.07&6 & &$-$0.06&0.11&9 & &$-$0.48&0.12&9 & &$-$0.41&0.09&11 \\
Ti\,{\sc i}  &$-$0.39&0.08&6 & &$-$0.03&0.13&17& &$-$0.51&0.11&20& &$-$0.23&0.14&24 \\
Ti\,{\sc ii} &       &    &  & &  0.07 &    &1 & &$-$0.48&    &1 & &$-$0.19&    &1  \\
V\,{\sc i}   &$-$0.44&0.08&6 & &$-$0.18&0.09&15& &$-$0.70&0.14&15& &$-$0.44&0.11&16 \\
Cr\,{\sc i}  &$-$0.36&0.12&2 & &$-$0.22&0.10&6 & &$-$0.87&0.13&6 & &$-$0.52&0.10&7  \\
Mn\,{\sc i}  &       &    &  & &$-$0.21&0.18&2 & &$-$0.89&0.04&2 & &$-$0.63&0.04&2  \\
Fe\,{\sc i}  &$-$0.48&0.05&24& &$-$0.18&0.08&29& &$-$0.76&0.08&27& &$-$0.55&0.12&39 \\
Fe\,{\sc ii} &$-$0.48&0.07&2 & &$-$0.18&0.08& 3& &$-$0.76&0.08&3 & &$-$0.55&0.10&5  \\
Co\,{\sc i}  &$-$0.41&0.16&3 & &$-$0.16&0.09&7 & &$-$0.67&0.12&7 & &$-$0.44&0.12&7  \\
Ni\,{\sc i}  &$-$0.42&0.15&9 & &$-$0.16&0.09&20& &$-$0.72&0.12&20& &$-$0.48&0.13&22 \\
Y\,{\sc i}   &       &    &  & &$-$0.24&    &1 & &       &    &  & &$-$0.47&    &1  \\
Y\,{\sc ii}  &       &    &  & &$-$0.13&0.11&3 & &$-$0.71&0.09&3 & &$-$0.48&0.05&2  \\
Zr\,{\sc i}  &$-$0.55&0.07&3 & &$-$0.22&0.09&4 & &$-$0.70&0.09&3 & &$-$0.32&0.06&3  \\
Ba\,{\sc ii} &$-$0.47&0.08&2 & &$-$0.14&0.09&2 & &$-$0.70&    &1 & &$-$0.43&0.09&2  \\
La\,{\sc ii} &$-$0.42&    &1 & &       &    &  & &       &    &  & &$-$0.48&    &1  \\
Sm\,{\sc ii} &       &    &  & &$-$0.01&    &1 & &$-$0.55&    &1 & &$-$0.41&    &1  \\
Eu\,{\sc ii} &       &    &  & &  0.00 &    &1 & &$-$0.20&    &1 & &$-$0.15&    &1  \\
             &       &    &  & &       &    &  & &       &    &  & &      &    &  \\
C/N          &       &    &  & & 1.05  &    &  & &  1.10 &    &  & &  0.85&    &   \\
$^{12}$C/$^{13}$C &  &    &  & & $>5$  &    &  & & 3     &+2/-1& & & $>5$ &    & \\
        \noalign{\smallskip}
   \cline{1-16}
         \end{tabular}
      \]
   \end{table*}


The theoretical standard stellar evolution of the surface carbon isotopic
ratios and carbon to nitrogen ratios along the giant branch was homogeneously
mapped by Charbonnel (1994) and more recently by Girardi et al. (2000) for
stellar masses between 1 and $7~M_{\odot}$ and different metallicities.
Our investigated field RHB stars are
somewhat metal deficient ($Z\approx0.008$) and have masses approximately
$0.8$ to $0.9~M_{\odot}$ (Tautvai\v{s}ien\.{e} 1996b).
According to Girardi et al. (2000), the C/N and $^{12}$C/$^{13}$C
ratios in such stars should drop after the first dredge-up episode
to values of about 3 and 35, respectively.
Charbonnel (1994, extrapolation to $\approx 0.85~M_\odot$ in
Figs.~2 and 4) predicted similar values after the first dredge-up.

It has long been known that giant stars regularly show much larger
evolutionary changes in these abundances than standard models predict,
see e.g. Boothroyd \& Sackmann (1999) for references.
This is the case also for our derived $^{12}$C/$^{13}$C and C/N ratios.
Because of grave differences between model predictions and observations,
Charbonnel (1995), Charbonnel et al. (1998) and Boothroyd \& Sackmann (1999)
performed calculations of models with deep mixing after the first dredge-up.
Boothroyd \& Sackmann e.g. fitted a one-parameter recipe for
``cool bottom processing'' (CBP) after the first dredge-up to the available
observations of red-giant abundances.
Their CBP results are given for initial stellar masses above $1.0~M_\odot$.
It is difficult to say what were the initial masses of the stars we investigate.
It could be that they lost about 0.1--0.3~$M_\odot$ during their evolution
on the giant branch (Renzini 1981; Renzini \& Fusi Pecci 1988).

The $^{12}\mathrm{C}/^{13}\mathrm{C}$ ratios determined for the investigated
stars are in quite good agreement
with ``cool bottom processing'' predictions (Boothroyd \&
Sackmann 1999) for low mass stars with $Z=0.007$.
The C/N ratios, however, request the initial mass of the stars to be of
about $1.8~M_\odot$. 
The metal-deficient RHB stars investigated by Gratton et al. (2000a) show 
higher than predicted by CBP $^{12}\mathrm{C}/^{13}\mathrm{C}$ ratios but 
even lower C/N ratios.
The low C/N ratios may be an indication that
CBP is stronger in such stars than the metallicity scaling of models
suggest. However, in view of the sensitivity of C/N ratios to the carbon
abundances,
we will not claim that the C/N predictions of Sackmann \& Boothroyd are
wrong, but rather that the C and N abundances should be checked in further
studies employing other atomic and molecular features.

\subsection{Sodium and aluminium}

   \begin{table}
  \setcounter{table}{3}
\centering
      \caption{(continued)}
      \[
         \begin{tabular}{lrcrc}
            \hline
            \noalign{\smallskip}
 & \multicolumn{3}{c}{HD 105944 } \\
            \noalign{\smallskip}
\cline{2-4}
            \noalign{\smallskip}
Ion & [A/H] &$\sigma$ &$n$\\
            \noalign{\smallskip}
            \hline
            \noalign{\smallskip}
C (C$_2$)    &$-$0.60&    &1 \\
N (CN)       &  0.07&0.13&77\\
O\,{\sc i}   &$-$0.29&    &1 \\
Na\,{\sc i}  &$-$0.24&    &1 \\
Mg\,{\sc i}  &$-$0.17&    &1 \\
Al\,{\sc i}  &$-$0.34&0.02&2 \\
Si\,{\sc i}  &$-$0.31&0.11&8 \\
Ca\,{\sc i}  &$-$0.19&0.12&6 \\
Sc\,{\sc i}  &$-$0.27&0.04&2 \\
Sc\,{\sc ii} &$-$0.28&0.11&9 \\
Ti\,{\sc i}  &$-$0.28&0.14&19\\
Ti\,{\sc ii} &$-$0.15&    &1 \\
V\,{\sc i}   &$-$0.36&0.11&14\\
Cr\,{\sc i}  &$-$0.41&0.14&7 \\
Mn\,{\sc i}  &$-$0.40&0.12&2 \\
Fe\,{\sc i}  &$-$0.37&0.08&30\\
Fe\,{\sc ii} &$-$0.37&0.06&3 \\
Co\,{\sc i}  &$-$0.38&0.11&8 \\
Ni\,{\sc i}  &$-$0.37&0.14&20\\
Y\,{\sc i}   &      &    &  \\
Y\,{\sc ii}  &$-$0.54&0.06&3 \\
Zr\,{\sc i}  &$-$0.29&0.10&2 \\
Ba\,{\sc ii} &$-$0.20&0.03&2 \\
La\,{\sc ii} &      &    &  \\
Sm\,{\sc ii} &$-$0.17&    &1 \\
Eu\,{\sc ii} &$-$0.16&    &1 \\
      &      &    &  \\
C/N  &   0.85&    &  \\
$^{12}\mathrm{C}/^{13}\mathrm{C}$ & 3.5 &+4/-2 & & \\
        \noalign{\smallskip}
   \cline{1-4}
         \end{tabular}
      \]
   \end{table}

Sodium and aluminium are among the mixing-sensitive chemical elements.
The star-to-star variations of Na, the existence of Na versus N correlations,
and Na versus O anticorrelations in globular cluster red giants have revealed
the possibility of sodium and aluminium are produced in red giant stars
(see Kraft 1994 and Da Costa 1998 for reviews).
It is found also that Na variations exist in
all clusters, while Al variations are greater in the more metal-poor clusters
(c.f. Norris \& Da Costa 1995, Shetrone 1996, Paper~I).

Pilachowski et al.\ (1996) determined sodium abundances for 60 metal-poor
halo subgiants, giants, and horizontal branch stars using high dispersion spectra
and concluded that there is an intrinsic difference between halo field
giants and globular cluster giants.
The bright giants in the field do not show the
sodium excesses seen in their globular cluster counterparts.
The [Na/Fe] ratios in field stars show a wide scatter (ranging from $-0.6$
to nearly $+0.3$)
with a slight tendency for $<$[Na/Fe]$>$ to increase with advancing
evolutionary stage.
In a sample of ten field RHB stars investigated by Tautvai\v{s}ien\.{e} (1997)
only two of the more metal rich ([Fe/H]$>-0.5$) stars showed
sodium overabundances of 0.2--0.3~dex.

The stars in our sample show Na and Al abundances which are typical of unevolved
stars in the solar vicinity,
as determined from the Na\,{\sc i} lines $\lambda$~5682.64 and
6154.23~{\AA} and Al\,{\sc i} lines
$\lambda$~6696.03, 6698.66, 7835.31 and 7836.13~{\AA}, see Fig.~5.
Gratton et al.\ (2000a) investigated possible non-LTE effects for the
Na\,{\sc i} lines, and find the probable corrections not to be larger than
about 0.02~dex at the temperatures and gravities of the stars analysed here.

Theoretical explanations for the production of Na and Al have been proposed
by Sweigart \& Mengel (1979), Langer \& Hoffman (1995), Cavallo et al.\ (1996),
Mowlavi (1999), Weiss et al.\ (2000) and other studies.
The nature and extent of the phenomenon is, however, still not well understood.

Prochaska et al.\ (2000) investigated abundances of Na and Al in 10 thick disk 
dwarfs and found aluminium to be much more overabundant than sodium. Our sample of 
thick disk stars does not show such a pattern. 
 
  \begin{figure}
    \resizebox{\hsize}{!}{\includegraphics{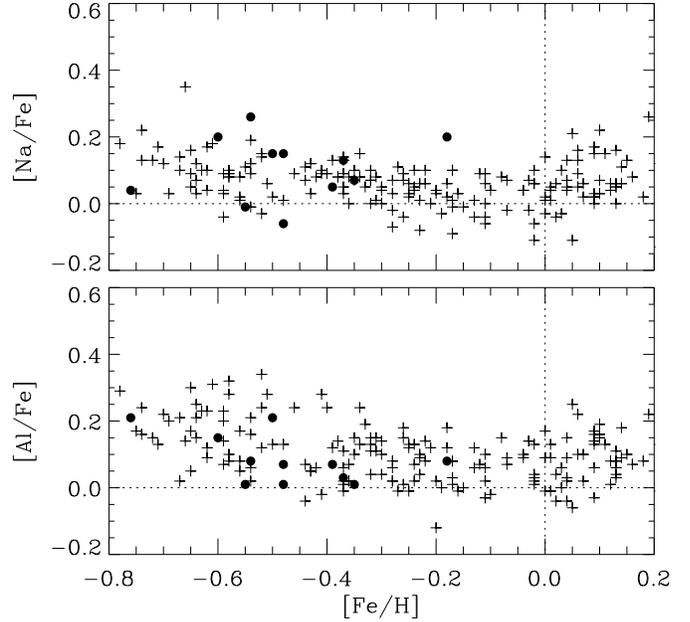}}
    \caption{[Na/Fe] and [Al/Fe] ratios as a function of iron [Fe/H].
Results for the field
RHB stars investigated in the present work are indicated by {\it filled circles},
for the Galactic disk stars investigated by Edvardsson et al.\ (1993)
by {\it crosses} }
\label{GT1664f5}
  \end{figure}

\subsection{Oxygen and magnesium}

Surface abundances of oxygen and magnesium could be altered in stars only by
very deep mixing. E.g., in cluster giants with large aluminium enhancements
($\sim 1.0$~dex) produced by very deep mixing, Mg depletions should then be
about $\sim 0.2$~dex (Langer \& Hoffman 1995). Since this is not the case for the
investigated stars we will discuss our results for oxygen and magnesium in
the context of the thick disk of the Galaxy.

In Figs.~6 and 7, we plot oxygen and magnesium abundance 
ratios and compare
them with the modeled ratios describing the mean trend of the Galactic thin disk
(Pagel \& Tautvai\v{s}ien\.{e} 1995). Other results obtained for the thick disk
stars in recent studies are displayed as well. Prochaska et al.\ (2000)
analysed a sample of 10 thick disk stars with the HIRES spectrograph on the
10~m Keck~I telescope. Unfortunately, the forbidden [O\,{\sc i}]
$\lambda 6300$~\AA\
line fell in the inter-order gap and the less trustworthy O\,{\sc i} triplet
lines at 7775~\AA\ had to be used in their analysis.
We adopt for the figures the results for 4 thick disk stars
from the work by Gratton et al.\ (2000b). In the same paper a sample
of thick disk candidates was selected from the work by Edvardsson et al.\
(1993). Stars which have [O/H]$>-0.5$, [Fe/O]$<-0.25$ and $-0.5<$[Mg/H]$<0$,
[Fe/Mg]$<-0.25$
and appropriate dynamical parameters were attributed to the
thick disk. While plotted, the data make quite a cloud lying above
the semiempirical trends modeled for the thin disk of the Galaxy by
Pagel \& Tautvai\v{s}ien\.{e} (1995), but this can hardly be used to draw any
conclusions about the location in terms of metallicity of the transition between
the halo and thick disk populations.
The high accuracy results for magnesium determined by Fuhrmann (1998) lie at the
edge of the distribution.
This may be taken as an indication that the transition between the halo phase
and the thick disk phase took place around
[Fe/H]\,$\approx -0.6$ to $-0.5$.
Our oxygen and magnesium to iron ratios tend to indicate the onset of
supernova of Type Ia (SN Ia) at about [Fe/H]$=-0.7$ to $-0.6$.
We suggest that a model for the halo and thick disk may look much like the
model of Pagel \& Tautvai\v{s}ien\.{e} (1995), with the difference that
the halo phase continued all the way up to [Fe/H]\,$\approx -0.6$~dex.

  \begin{figure}
    \resizebox{\hsize}{!}{\includegraphics{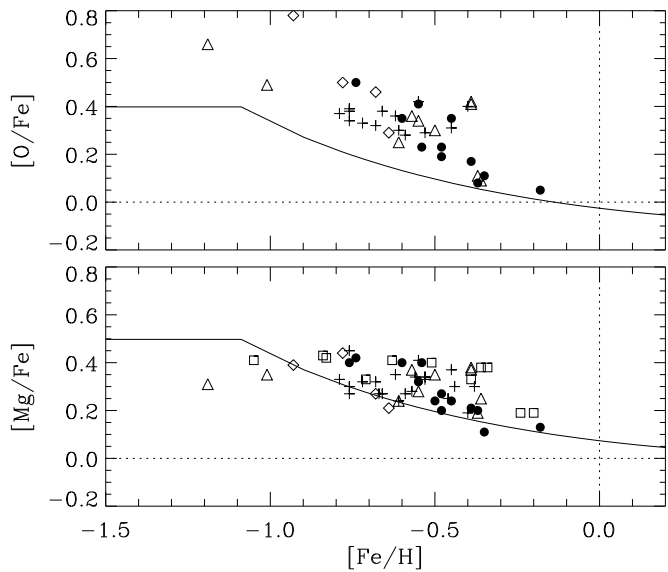}}
    \caption{[O/Fe] and [Mg/Fe] ratios as a function of iron [Fe/H]
    for the thick disk stars analysed in recent studies:
     {\it filled circles} -- the present work;
{\it triangles} -- Prochaska et al.\ (2000);
{\it rhombs} -- Gratton et al.\ (2000b);
{\it crosses} -- Edvardsson's et al.\ (1993) dwarfs with $R_m\le 7$kpc,
reanalysed and selected to be the thick disk stars by Gratton et al.\ 2000b;
{\it squares} -- Fuhrmann (1998).
The solid lines show the model of the Galactic thin disk (Pagel \&
Tautvai\v{s}ien\.{e} 1995) }
\label{GT1664f6}
  \end{figure}

  \begin{figure}
    \resizebox{\hsize}{!}{\includegraphics{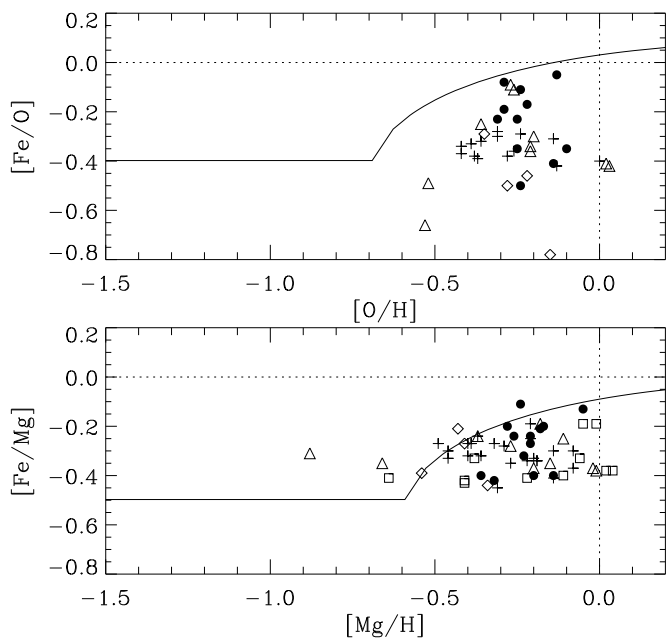}}
    \caption{Run of [Fe/O] vs. [O/H] and [Fe/Mg] vs. [Mg/H] ratios
    for the stars of Figure~6 }
    \label{GT1664f7}
  \end{figure}

\subsection{Silicon, calcium and titanium}

The $\alpha$-elements silicon, calcium and titanium may also bring information
on the thick disk of the Galaxy. A large number of spectral lines with accurate
$gf$-values are available for the analysis which should provide for good
abundance precision.
Being produced both in Type II and Ia supernova, Si, Ti and Ca may be expected
to show smaller overabundances than O and Mg.
As is seen from Fig.~8, abundance ratios of these elements to iron may also
exhibit slight overabundances with respect to the mean trend of the thin disk.

  \begin{figure}
    \resizebox{\hsize}{!}{\includegraphics{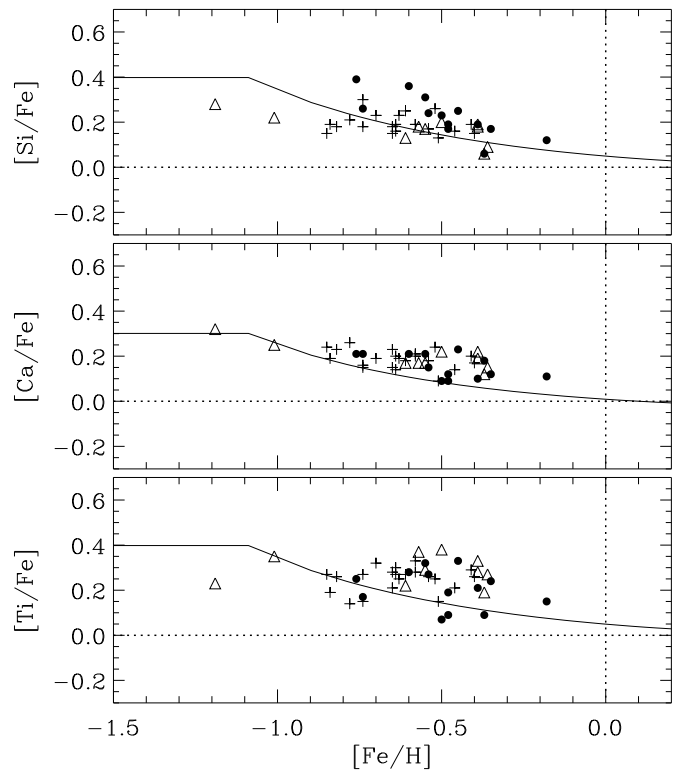}}
    \caption{[Si/Fe], [Ca/Fe] and [Ti/Fe] ratios as a function of iron [Fe/H]
    for the thick disk stars analysed in recent studies. The meaning
    of symbols as in Fig.~6} 
    \label{GT1664f8}
  \end{figure}

\subsection{$s$- and $r$-process elements}

As already mentioned, the barium abundances
in our study are corrected for non-LTE effects by the subtraction of 0.20~dex.
Two quite similar Ba\,{\sc ii} lines $\lambda 6141$ and $6496 $~\AA\ were used
for the analysis.
According to Mashonkina et al.\ (1999) and Mashonkina \& Gehren (2000),
the non-LTE correction for the Ba\,{\sc ii} line $\lambda 6496$ is $-0.2$~dex on
average in the metallicity range $-1 <$[Fe/H]$<0.1$.
Non-LTE effects for the line $\lambda 6141$ were not
studied well enough, since this line is too saturated in the solar spectrum
to provide an accurate correction. Theoretical non-LTE calculations
show that non-LTE effects for this line are not smaller than for
$\lambda 6496$, only the weak line $\lambda 5853$~\AA\ is quite insensitive.
In our study, both $\lambda 6141$ and 6496~\AA\ gave approximately the
same barium abundances, so we decided to apply the same correction to both.
In the work by Prochaska et al.\ (2000)
three Ba\,{\sc ii} lines $\lambda 5853$, 6141 and 6496 were used, and a typical
correction of 0.17~dex was applied.

  \begin{figure}
    \resizebox{\hsize}{!}{\includegraphics{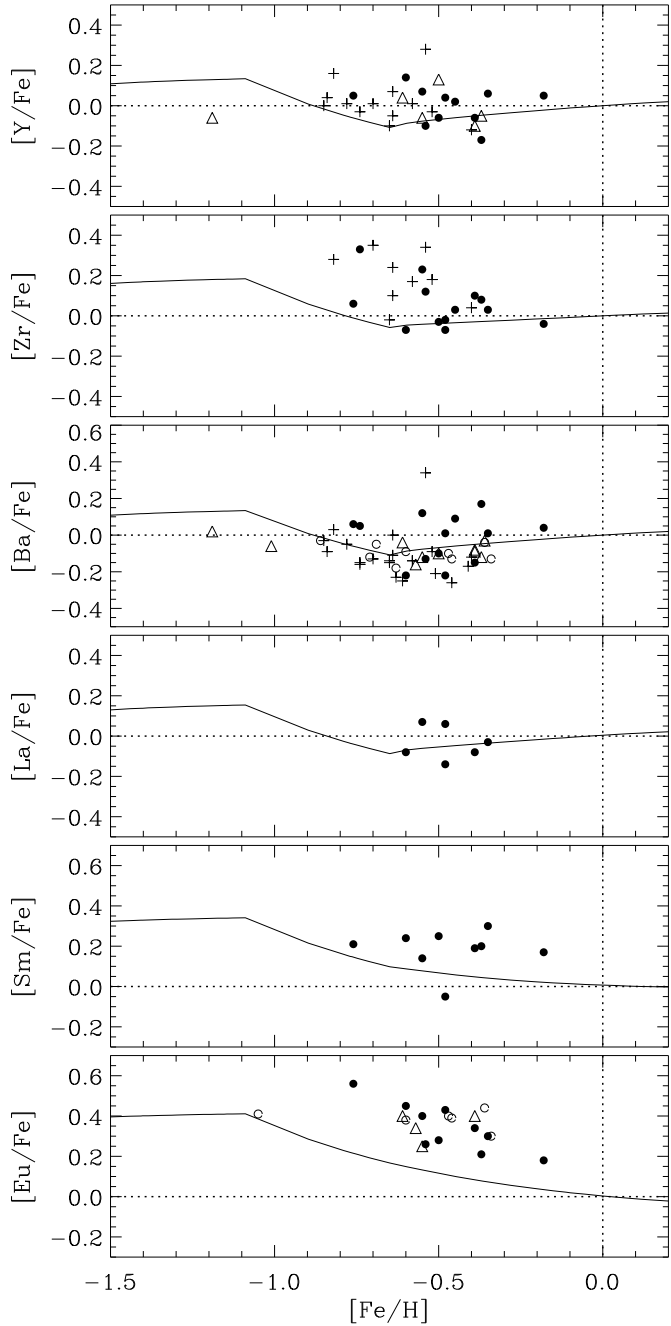}}
    \caption{Abundance ratios of the $s$-process dominated (Y, Zr, Ba and La)
    and $r$-process dominated (Sm and Eu) elements to iron as a
    function of iron [Fe/H]
    for the thick disk stars analysed in recent studies. The meaning of
    symbols as in Fig.~6, {\it open circles} represent results by
    Mashonkina \& Gehren (2000). The solid line shows the model of the
    Galactic thin disk (Pagel \& Tautvai\v{s}ien\.{e} 1997) }
    \label{GT1664f9}
  \end{figure}

Abundance ratios of $s$- and $r$-process-dominated (in the Solar system,
Burris et al.\ 2000) elements to iron as a function of
iron [Fe/H] for the thick disk stars analysed in the recent studies are
presented in Fig.~9. 
For a comparison, the modeled abundance trends
of the Galactic thin disk by Pagel \& Tautvai\v{s}ien\.{e} (1997) are shown.
As is the case for oxygen and the $\alpha$ elements,
these elements fit the models for the thin disk reasonably well
if we shift the onset of SN Ia from [Fe/H]$=-1.1$ to $-0.6$~dex.
Since europium is an almost pure $r$-process element and supposedly
produced with oxygen and magnesium in stars exploding as
core-collapse supernovae, the thick-disk Eu abundance trend differ quite 
dramatically from the thin-disk one and may be very useful for population 
studies. [Eu/Fe] ratios obtained in our sample of thick disk stars and in ten 
more stars analysed by Prochaska et al.\ (2000) and Mashonkina \& Gehren 
(2000) bring quite a clear indication that
the thick disk population is chemically discrete from the thin disk.

\section{Summary and conclusions}

More than ten years have passed since the high-resolution spectroscopic study
by Barbuy \& Erdelyi-Mendes (1989), in which a spread in [O/Fe] ratios at
$-0.8<$[Fe/H]$<-0.5$ was proposed to be an indication of a thick disk phase in
the chemico-dynamic evolution of the Galaxy. However, further studies
by observers and theoreticians did not bring accurate enough characterization
of the thick disk of the Galaxy (c.f. Pagel 2001, Bernkopf et al.\ 2001,
Nissen 1999, Chiappini et al.\ 1997, Robin et al.\ 1996). The thick disk
still needs to be revisited by new observations.

We have presented a detailed chemical abundance analysis of 13 core
helium-burning low-mass stars, representatives of the thick disk of the Galaxy.
Abundances in the investigated stars show that carbon is depleted by
about 0.3~dex, nitrogen is enhanced by more than 0.4~dex,
the $^{12}\mathrm{C}/^{13}\mathrm{C}$
ratios are lowered to values from 3 to 7 and C/N ratios to values from 0.7
to 1.2.
These abundance ratios can only be accounted for by stellar evolution
calculations if extra mixing, e.g. ``cool bottom processing'' (Boothroyd \&
Sackmann 1999), after the first dredge-up episode is prescribed.

In agreement with other studies of field core-helium-burning stars, our stars
do not show enhanced overabundances of Na and Al.

Abundance ratios of O, Mg, Eu and other heavy elements to iron in the
investigated stars provide further evidence that the thick disk population had a
different chemical history as compared to the thin disk (c.f. Fuhrmann 1998, 
Gratton et al.\ 2000b, Prochaska et al.\ 2000).
We propose that the time-scale for metal enrichment was short
for the thick disk population, and thet SN Ia started to contribute
with iron-peak nuclei only after the overall metallicity reached
[Fe/H]$\approx -0.7$ or $-0.6$~dex.

\begin{acknowledgements}
We wish to acknowledge B.E.J. Pagel and A.I. Boothroyd for insightful
discussion and comments.
Heidi Korhonen (NOT) and Eduaras Puzeras (ITPA) are thanked for their help
in spectral reductions.
Bertrand Plez (University of Montpellier II) and
Guillermo Gonzalez (Washington State University) were particularly generous
in providing us with atomic data for CN and C$_2$ molecules, respectively.
We are very grateful to Martin Asplund (Uppsala Astronomical Observatory)
for computing of the necessary stellar model atmospheres. We also thank 
the referee, J.A. Rose, for valuable comments on the manuscript. 
This research has made use of Simbad and VALD databases.
G.T. acknowledges support from NATO Linkage grant CRG.LG 972172.
B.E. was supported by the Swedish Natural Sciences Research Council (NFR). 
I.T. and I.I. acknowledge the Academy of Finland for the research grants 
44153 and 10848. 
\end{acknowledgements}


\begin{thebibliography}{}
    \bibitem{} Asplund M., Gustafsson B., Kiselman D., Eriksson K. 1997,
A\&A 318, 521
    \bibitem{} Barbuy B., Erdelyi-Mendes M. 1989, A\&A 214, 239
    \bibitem{} Bartkevi\v{c}ius A., Lazauskait\.{e} R. 1997, Baltic Astron. 
6, 499 
    \bibitem{} Bernkopf J., Fiedler A., Fuhrmann K. 2001, Nature (in press)
    \bibitem{} Boothroyd A.I., Sackmann I.-J. 1999, ApJ 510, 232
    \bibitem{} Burris D.L., Pilachowski C.A., Armandroff T.E., Sneden C.,
Cowan J.J., Roe, H. 2000, ApJ 544, 302
    \bibitem{} Cavallo R.M., Sweigart A.V., Bell R.A. 1996, ApJ 464, L79
    \bibitem{} Charbonnel C. 1994, A\&A 282, 811
    \bibitem{} Charbonnel C. 1995, ApJ 454, L41
    \bibitem{} Charbonnel C., Brown J.A., Wallerstein G. 1998, A\&A 332, 204
    \bibitem{} Chiappini C., Matteucci F., Gratton R. 1997, ApJ 477, 765
    \bibitem{} Clegg R.E.S., Lambert D.L., Tomkin J. 1981, ApJ 250, 262
    \bibitem{} Da Costa G.S. 1998, IAUS 189, 193
    \bibitem{} Edvardsson B., Andersen J., Gustafsson B., Lambert D.L.,
Nissen P.E., Tomkin J. 1993, A\&A 275, 101
    \bibitem{} Fuhrmann K. 1998, A\&A 338, 161
    \bibitem{} Gilmore G., Reid N. 1983, MNRAS 202, 1025
    \bibitem{} Girardi L., Bressan A., Bertelli G., Chiosi C. 2000,
A\&AS 141, 371
    \bibitem{} Gonzalez G., Lambert D.L., Wallerstein G., et al. 1998, ApJS
114, 133
    \bibitem{} Gratton R.G., Sneden C., Carretta E., Bragaglia A. 2000a,
A\&A 354, 169
    \bibitem{} Gratton R.G., Carretta E., Matteucci F., Sneden C. 2000b,
A\&A 358, 670
    \bibitem{} Gurtovenko E.A., Kostik R.I. 1989, Fraunhofer's spectrum and
a system of solar oscillator strengths, Kiev, Naukova Dumka, p. 200
    \bibitem{} Gustafsson B., Bell R.A., Eriksson K., Nordlund {\AA} 1975,
A\&A 42, 407
    \bibitem{} Gustafsson B., Karlsson T., Olsson E., Edvardsson B., Ryde N.
1999, A\&A 342, 426
    \bibitem{} Ilyin I.V. 1996, Remote control observation with the SOFIN
spectrograph and reduction of CCD \'{e}chelle spectra. Licentiate
dissertation, Univ. Oulu, Finland
    \bibitem{} Ilyin I.V. 2000, High resolution SOFIN CCD \'{e}chelle 
spectroscopy, PhD dissertation, Univ. Oulu, Finland
    \bibitem{} Kraft R.P. 1994, PASP 106, 553
    \bibitem{} Kurucz R.L., Furenlind I., Brault J., Testerman L. 1984,
Solar Flux Atlas from 269 to 1300 nm, National Solar Observatory, Sunspot,
New Mexico
    \bibitem{} Langer G.E., Hoffman R.D. 1995, PASP 107, 1177
    \bibitem{} Mackay C.D. 1986, ARA\&A 24, 255
    \bibitem{} Mashonkina I., Gehren T. 2000, A\&A 364, 249
    \bibitem{} Mashonkina I., Gehren T., Bikmaev I.F. 1999, A\&A 343, 519
    \bibitem{} Moore C.E., Minnaert M.G.J., Houtgast J. 1966, The Solar
Spectrum 2935~\AA\ to 8770~\AA\, NBS Monogr., No.~61
    \bibitem{} Mowlavi N. 1999, A\&A 350, 73
    \bibitem{} Nissen P.E. 1999, Ap\&SS 265, 249
    \bibitem{} Norris J. 1987, AJ 93, 616
    \bibitem{} Norris J.E., Da Costa G.S. 1995, ApJ 441, L81
    \bibitem{} Pagel B.E.J. 2001, in Cosmic Evolution, eds. E. Vangioni-Flam
\& M. Cass\'{e}, Paris IAp Coll., World Scientific (in press)
    \bibitem{} Pagel B.E.J., Tautvai\v{s}ien\.{e} 1995, MNRAS 276, 505
    \bibitem{} Pagel B.E.J., Tautvai\v{s}ien\.{e} 1997, MNRAS 288, 108
    \bibitem{} Pilachowski C.A., Sneden C., Kraft R.P. 1996, AJ 111, 1689
    \bibitem{} Piskunov N.E., Kupka F., Ryabchikova T.A., Weiss W.W.,
Jeffery C.S. 1995, A\&AS 112, 525
    \bibitem{} Plez B. 1999, private communication
    \bibitem{} Prieto C.A., Lambert D.L., Asplund M. 2001, astro-ph/0106360
    \bibitem{} Prochaska J.X., Naumov S.O., Carney B.W., McWilliam A.,
Wolfe A.M. 2000, AJ 120, 2513
    \bibitem{} Renzini A. 1981, in Effects of Mass Loss on Stellar Evolution,
ed. C. Chiosi \& R. Stalio (Dordrecht:\,Reidel), 319
    \bibitem{} Renzini A., Fusi Pecci F. 1988, ARA\&A, 26, 199
    \bibitem{} Robin A., Haywood M., Cr\'{e}z\'{e} M., Ojha D.K., Bienayme O.
1996, A\&A 305, 125
    \bibitem{} Rose J. A. 1985, AJ 90, 787
    \bibitem{} Shetrone M.D. 1996, AJ 112, 1517
    \bibitem{} Stetson P.B., Aikman G.Ch. 1987, AJ 93, 1439
    \bibitem{} Sturch C., Helfer H.L. 1971, AJ 76, 334
    \bibitem{} Sweigart A.V., Mengel J.G. 1979, ApJ 229, 624
    \bibitem{} Tautvai\v{s}ien\.{e} G. 1996a, Baltic Astron. 5, 503 
    \bibitem{} Tautvai\v{s}ien\.{e} G. 1996b, Astron. Nachr. 317, 29
    \bibitem{} Tautvai\v{s}ien\.{e} G. 1997, MNRAS 286, 948
    \bibitem{} Tautvai\v{s}ien\.{e} G., Edvardsson B., Tuominen I., Ilyin I.
2000, A\&A 360, 499, (Paper I)
    \bibitem{} Tuominen I. 1992, NOT News, No. 5, p. 15
    \bibitem{} Upgren A.R. 1962, AJ 67, 37
    \bibitem{} Weiss A., Denissenkov P.A., Charbonnel C. 2000, A\&A 356, 181
    \bibitem{} Wyller A.A. 1966, ApJ 143, 828.
\end{thebibliography}
\end{document}